\documentclass[aps,prl,twocolumn,showpacs,superscriptaddress,nofootinbib]
{revtex4-1}

\usepackage{amssymb}
\usepackage{graphicx}

\begin{document}

\title{Electromagnetic prompt response in an elastic wave cavity}

\author{A. M. Mart\'inez-Arg\"uello}
\author{M. Mart\'inez-Mares}
\affiliation{Departamento de F\'{\i}sica, Universidad Aut\'onoma
Metropolitana-Iztapalapa, Apartado Postal 55-534, 09340 M\'exico Distrito
Federal, Mexico}

\author{M. Cobi\'an-Su\'arez} 
\author{G. B\'aez} 
\affiliation{Departamento de Ciencias B\'asicas, Universidad Aut\'onoma
Metropolitana-Azcapotzalco, Apartado Postal 21-267, 04000 M\'exico Distrito
Federal, Mexico}

\author{R. A. M\'endez-S\'anchez}
\affiliation{Instituto de Ciencias F\'isicas, Universidad Nacional Aut\'onoma de
M\'exico, Apartado Postal 48-3, 62210 Cuernavaca Mor., Mexico}

\begin{abstract}
A  rapid, or prompt response, of an electromagnetic nature, is found in an elastic wave 
scattering experiment. The experiment is performed with torsional elastic waves 
in a quasi-one-dimensional cavity with one port, formed by a notch grooved at a 
certain distance from the free end of a beam. The stationary patterns are 
diminished using a passive vibration isolation system at the other end of the 
beam. The measurement of the resonances is performed with non-contact 
electromagnetic-acoustic transducers outside the cavity. In the Argand plane, 
each resonance describes a circle over a base impedance curve which comes from 
the electromagnetic components of the equipment. A model, based on a variation 
of Poisson's kernel is developed. Excellent agreement between theory and 
experiment is obtained.
\end{abstract}


\maketitle

The study of wave transport through cavities, or scattering properties in a more 
general sense, has received intense theoretical and experimental attention in 
recent years. On the one hand, special interest has been the fingerprint of the 
classical dynamics in wave 
systems~\cite{MelloKumar,MelloBaranger,Beenakker,Alhassid}, 
like lithographic quantum dots~\cite{Marcus,Chang,Chan,Keller}, microwave
cavities~\cite{MendezSanchez,Kuhl,Schanze,Doron,Graf,Alex} and
graphs~\cite{Sirko1,Sirko2}. On the other hand, most of the experiments on 
elastic systems with different geometries are mainly concerned with the analysis 
of the spectrum~\cite{Weaver,Schaadt,Mori,Arreola-LucasFranco-VillafaneBaezMendez-Sanchez,Xeridat} or carried out in 
the time domain~\cite{Fink,EPL2011}. Single-frequency scattering experiments on 
elastic systems are difficult to perform due to the formation of stationary 
patterns~\cite{LobkisRozhkovWeaver}. Also, the excitation and selection of a 
particular type of elastic motion is a difficult task with commercial 
piezoelectric transducers.

Recently, this has started to change due to the successful reduction of the stationary waves 
using absorbing materials~\cite{Xeridat,APP}. 
In adition, particular kinds of vibrations can be selectively excited or detected with 
different configurations of electromagnetic-acoustic transducers 
(EMATs)~\cite{Mori,refEMAT,Arreola-LucasFranco-VillafaneBaezMendez-Sanchez}. 
Although the use of EMATs provide the advantage of not being in contact with 
the system, the electromagnetic field of the exciter enters into the 
detector~\cite{APP}. Therefore, the elastic scattering system as a whole, 
including the EMATs, presents an electromagnetic impedance which cannot be 
ignored. The resonances of the system lie on a base electromagnetic impedance curve that induces direct processes comming from an electromagnetic prompt response in the scattering of elastic waves, as well as the expected prompt response in the elastic waves. As will be shown below, the effect of this impedance can be taken into account in the total prompt response of the system.\footnote{Notice that this effect is not the same as that of Ref.~\cite{Hemmady1} where direct reflections 
are induced by imperfect coupling to the system.}

In this Letter we present measurements of the $1\times 1$ scattering matrix, in 
a single port quasi-one-dimensional cavity, for torsional waves in a beam. In 
the Argand plane, the experimental scattering matrix moves along the impedance 
curve describing circles on each resonance of the cavity, as frequency is 
varied. That is, the measured scattering matrix consists of the scattering 
matrix of the cavity plus its displacement along that impedance curve. 
Here, the experimental results are described by taking into account this phenomenon in a theoretical model based on Poisson's kernel, an opposite to that in which the direct processes are subtracted in microwave measurements in cavities~\cite{Hemmady1} and graphs~\cite{Sirko1}.

\begin{figure}
\includegraphics[width=1.0\columnwidth]{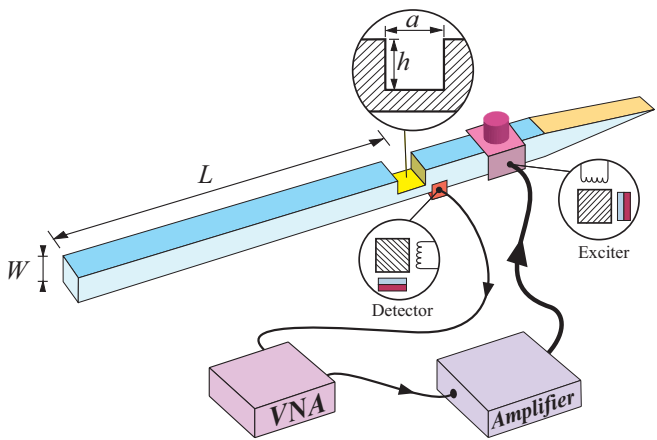}
\caption{Experimental setup used to form an open elastic wave system (not to scale).
A cavity is formed in a beam by machining a notch in it at a certain distance from
the free end. The total length of the beam is $3.6$~m with a square 
cross-section
of side $W=25.4$~mm; $L=2.5$~m, $h=18.0$~mm, and $a=0.9$~mm. The wedge is
$40.0$~cm long. The beam is supported by two nylon threads (not shown). The 
connection of the equipment is shown at the bottom. The exciter is located at 
70~cm from the notch. Aluminum alloy 6061-T6, with shear modulus $G=26$~GPa 
and density $\rho=2.7$~g/cm$^3$, was used.}
\label{fig:ExperimentalSetup}
\end{figure}

The cavity is formed on a square-cross-section aluminum beam, as shown in 
Fig.~\ref{fig:ExperimentalSetup}, between a free end and a notch of width $a$ 
and depth $h$, grooved at a distance $L$ from the free end. At the other end of 
the beam a passive vibration isolation system is used. Torsional 
vibrations are induced using electromagnetic-acoustic transducers (EMATs). The 
signal of a vector network analyzer (VNA, Anritsu MS-4630B) is sent to a 
high-fidelity audio amplifier (Cerwin-Vega CV-900) and then to the EMAT exciter 
located at a certain distance from the notch, outside the cavity. The torsional 
acceleration is measured by another EMAT, just outside the cavity, and sent to 
the VNA. In this way the non-normalized $1\times 1$ scattering matrix,
\begin{equation}
S(f) = \sqrt{R(f)}\, \mathrm{e}^{\mathrm{i}\theta(f)}, 
\end{equation}
where $R(f)$ is the reflection coefficient and $\theta(f)$ the phase, is 
measured for torsional waves. 

\begin{figure}
\includegraphics[width=1.0\columnwidth]{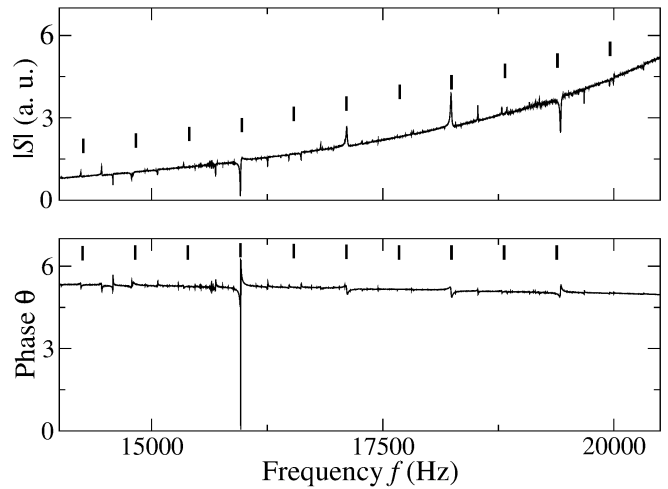}
\caption{The magnitude and phase of $S(f)$, showing the resonance
spectrum measured with the detector located just outside the cavity. The
vertical marks correspond to theoretical predictions~\cite{APP}.}
\label{fig:Spectrum}
\end{figure}

\begin{figure}
\includegraphics[width=0.8\columnwidth]{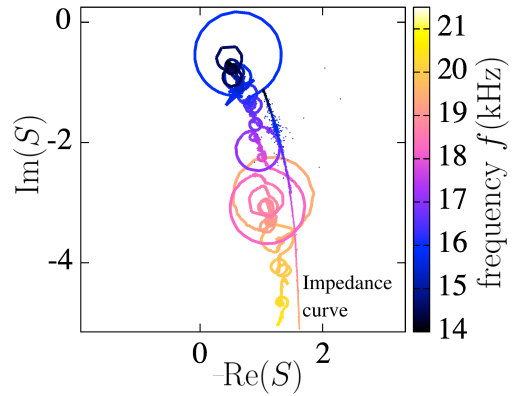}
\caption{(Color online) Motion of $S(f)$ in the Argand plane as a function of
frequency $f$ for the same resonances of Fig.~\ref{fig:Spectrum}. The impedance 
curve, measured without the elastic system, is given by the thick line.}
\label{fig:motion}
\end{figure}

A typical measurement is shown in Fig.~\ref{fig:Spectrum}, in which the magnitude 
and phase of $S$ are plotted as a function of the frequency $f$, from 14000 to 
20500~Hz. According to a simple theoretical model~\cite{APP}, eleven torsional resonances 
can be identified in this range of frequencies; they are 
indicated with vertical marks in the same figure. The remaining peaks correspond 
to other modes of vibration or to radio broadcasting stations~\cite{APP}. It is 
noticeable that all resonances are located on a base line which comes from the 
electromagnetic impedance of the system as a whole (see also 
Fig.~\ref{fig:motion}). This was verified by the null hypothesis that 
corresponds to the measurement without the elastic wave system. 

In Fig.~\ref{fig:motion} the motion of $S(f)$ in the complex plane is shown 
as a function of the frequency $f$. Instead of calibrating the scattering 
matrix as in Refs.~\cite{Hemmady1,Sirko1}, an unnormalized $S$-matrix will be 
used and the experimental data will be given in arbitrary units. As expected, 
$S(f)$ describes a circle for each resonance, but their centers are not located at 
the origin. In fact, the circles are superimposed on a base curve which is 
parallel to that obtained from the null hypothesis. The circles are not closed because the electromagnetic impedance is not constant but it varies slowly with frequency. 
Due to this fact each resonance is analyzed separately. 
In Fig.~\ref{fig:motion} the result of the null hypothesis is also shown in the Argand plane as the frequency is varied. 

As has been recently shown by the authors~\cite{APP,Angel}, 
the scattering matrix of each resonance, as seen 
from the center of its own circle, $S'(f)$, visits the circumference in a non-uniform way, {\em i.e.}, according to Poisson's kernel 
\begin{equation}
\label{eq:distsp}
P'(S') = \frac{1}{2\pi} 
\frac{R'-\left|\overline{S'}\right|^2}
{\left|S'-\overline{S'}\right|^2}\, 
\delta(R'-R_0),
\end{equation}
where 
\begin{equation}
S'(f)=\sqrt{R'}\,\mathrm{e}^{\mathrm{i}\theta'(f)}, \quad R'=R_0,
\end{equation}
and $R_0$ is a constant, experimentally determined as the square of the average
of $|S'(f)|$. Here, $\overline{S'}\equiv\overline{S'(f)}$ is the 
frequency average of $S'(f)$, also determined experimentally. In the jargon 
of nuclear physics this corresponds to the well known \emph{optical scattering 
matrix} which accounts for the prompt response of the system during the scattering 
processes. For the theoretical description of the experimental data, as 
originally measured, the effect of the electromagnetic impedance has to be taken 
into account. 


From the known distribution of $S'(f)$, Eq.~(\ref{eq:distsp}), the probability
distribution of $S(f)$ for each resonance can be obtained. Both scattering
matrices are related by a translation,
\begin{equation}
S(f) = S'(f) + Z, 
\end{equation}
where $Z$ is the position of the center of the circle, as seen from the origin; 
this displacement is triggered by the impedance, which is different for each 
resonance. This implies that the optical $S$ matrix, 
$S_{\rm{opt}}\equiv\overline{S(f)}$, consists of two parts: one coming from the 
scattering process of the system itself and a second coming from 
the electromagnetic components;
\begin{equation}
S_{\rm{opt}} = \overline{S'} + Z.
\end{equation}

From Eq.~(\ref{eq:distsp}), the probability density distribution of 
$S(f)$ is given by
\begin{equation}
\label{eq:dists}
P(S) = \frac{1}{2\pi}\, 
\frac{\left| S - Z \right|^2- 
\left|S_{\rm{opt}} - Z\right|^2} {\left| S - S_{\rm{opt}} \right|^2}\,
\delta[R'(S) - R_0], 
\end{equation}
where $R'(S)=|S-Z|^2$. In contrast to the ideal case where the impedance is
absent and only $S_{\rm{opt}}$ is enough to describe the scattering in the
system~\cite{Lopez,MelloSeligmanPereyra}, this distribution depends
on two experimental parameters, $S_{\rm{opt}}$ and $Z$. $R_0$ is a constant,
not a parameter, whose value deviates from 1 due to the arbitrary units used in the
measurements.  

\begin{figure}
\includegraphics[width=1.0\columnwidth]{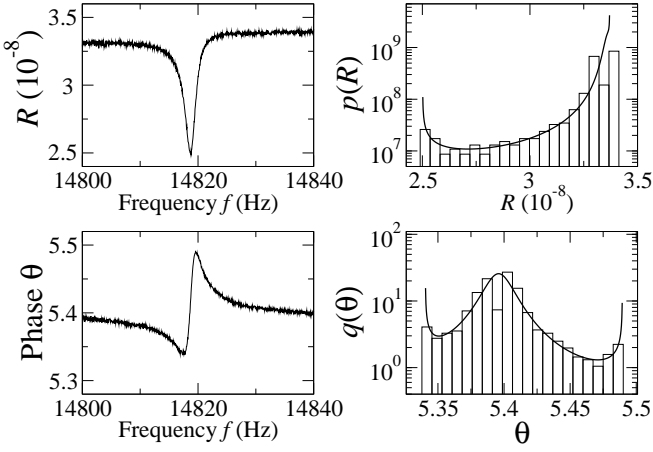}
\caption{Resonance at 14819~Hz. In the left panels the magnitude squared
and phase of $S(f)$ are shown. The histograms of the experimental data are compared with the
theoretical expressions (continuous lines) of Eqs.~(\ref{eq:p(R)}) and (\ref{eq:q(q)}).}
\label{fig:14800}
\end{figure}

To visualize the distribution of $S$, the marginal distributions for $R$ and 
$\theta$ will be determined. To obtain the distribution of $R$, 
Eq.~(\ref{eq:distsp}) is integrated with respect to $\theta$. It is clear that 
the minimum and a maximum values of $\sqrt{R}$ are imposed by $Z$; 
that is, the minimum (maximum) is given by the sum of the distance to center 
$|Z|$, minus (plus) the radius of the circle $\sqrt{R_0}$. From 
Eq.~(\ref{eq:dists}) we obtain 
\begin{eqnarray}
p(R) & = & \frac{1}{2\pi} 
\frac{R_0 - |S_{\rm{opt}}-Z|^2}
{\sqrt{ \big[R- (|Z|-\sqrt{R_0} )^2\big]
\big[ (|Z|+\sqrt{R_0} )^2-R\big] }} 
\nonumber \\ & \times & 
\left( \frac{1}{| \sqrt{R}\, \mathrm{e}^{\mathrm{i}\theta_{+}}-
S_{\rm{opt}} |^2} + 
\frac{1}{| \sqrt{R}\, \mathrm{e}^{\mathrm{i}\theta_{-}}- 
S_{\rm{opt}} |^2} \right) \label{eq:p(R)}
 \\ & \times & [
\Theta(|Z|+\sqrt{R_0}-\sqrt{R}) + 
\Theta(\sqrt{R}-|Z|+\sqrt{R_0})], \nonumber
\end{eqnarray}
where $\Theta(z)$ is the Heaviside function of $z$ and
\begin{eqnarray}
& & \sqrt{R}\, \mathrm{e}^{\mathrm{i}\theta_{\pm}} = 
\frac{1}{2Z^*} 
\bigg\{ (R + |Z|^2- R_0) \nonumber \\ & & \pm 
\mathrm{i}
\sqrt{ \big[R- (|Z|-\sqrt{R_0} )^2\big]
\big[ (|Z|+\sqrt{R_0} )^2-R\big] }
\bigg\}.\,
\end{eqnarray}

\begin{figure}
\includegraphics[width=1.0\columnwidth]{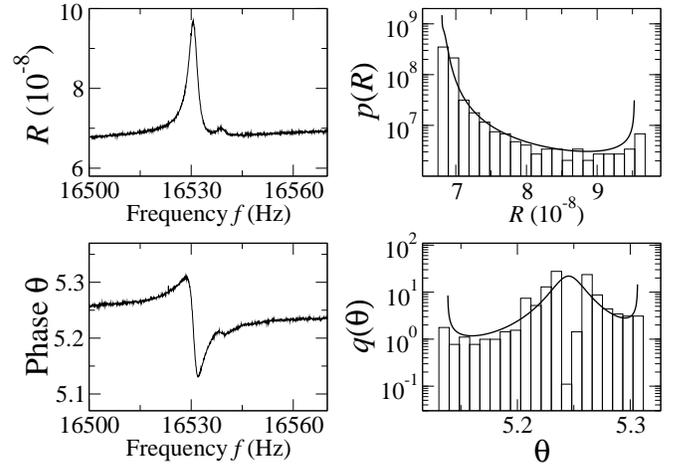}
\caption{Resonance at 16530~Hz. Excellent agreement is found between theory and
experiment despite the appearance of a second type of resonance, not completely eliminated by
the absorber.}
\label{fig:16500}
\end{figure}

In a similar way, the marginal phase distribution $q(\theta)$ is obtained 
by integrating over $R$; the result is
\begin{eqnarray}
& & q(\theta) = \frac{1}{2\pi} 
\frac{R_0 - |S_{\rm{opt}}-Z|^2}
{\sqrt{R_0-\big[\mathrm{Im}\big(Z^*\mathrm{e}^{\mathrm{i}\theta}\big)\big]^2 }} 
\nonumber \\ & & \times \left( 
\frac{\sqrt{R_+}}{|\sqrt{R_+}\, \mathrm{e}^{\mathrm{i}\theta}-
S_{\rm{opt}}|^2 } + 
\frac{\sqrt{R_-}}{|\sqrt{R_-}\, \mathrm{e}^{\mathrm{i}\theta}-
S_{\rm{opt}}|^2 }\right) \label{eq:q(q)} 
\\ & & \times \Big[
\Theta\bigg(\phi-\sin^{-1}\frac{\sqrt{R_0}}{|Z|} - \theta\Big) + 
\Theta\Big(\theta-\phi-\sin^{-1}\frac{\sqrt{R_0}}{|Z|}\Big)\bigg],
\nonumber
\end{eqnarray}
where $\phi$ is defined through $\mathrm{e}^{\mathrm{i}\phi}=Z/|Z|$; 
\begin{equation}
\sqrt{R_{\pm}} = \mathrm{Re}\big(Z^*\mathrm{e}^{\mathrm{i}\theta}\big) 
\pm \sqrt{R_0-\big[\mathrm{Im}
\big(Z^*\mathrm{e}^{\mathrm{i}\theta}\big)\big]^2 },
\end{equation}
and $\mathrm{Re}(z)$ and $\mathrm{Im}(z)$ stand for the real and imaginary
parts of $z$.

In Figs.~\ref{fig:14800} and \ref{fig:16500} the results for two typical 
experimental resonances, at 14819 and 16530~Hz, are shown. 
The left panels of both figures show the magnitude squared ($R$) and phase
($\theta$) of the amplitude of the measured signal; the effect of the 
impedance is noticeable. In the right panels the distributions for $R$ and 
$\theta$ are shown as histograms along with the theoretical expressions 
(continuous lines) for the marginal distributions, Eqs.~(\ref{eq:p(R)}) and 
(\ref{eq:q(q)}). The only relevant parameters, 
$S_{\rm{opt}}$ and $Z$, were obtained from the experimental data. Despite the 
hollow that appears in the histograms of the phase, an excellent agreement is 
observed. The hollow is due to the lack of data since the experimental resonances 
does not close completely. In Fig.~\ref{fig:16500} one can observe that 
another resonance appears around 16540~Hz. The effect of this resonance, that 
belongs to another type of vibration not completely eliminated by the passive 
vibration isolation system, does not noticeably affect the distributions.


In conclusion, wave transport through a quasi-one-dimensional elastic cavity 
was studied. This was done by measuring the scattering matrix for torsional 
waves. The system was opened at one end by a passive vibration isolation system 
reducing substantially the presence of stationary waves. Although the excitation (and 
detection) of the modes was performed by non-contact electromagnetic-acoustic 
transducers, the exciter induces an electromagnetic impedance in the detector 
that affects dramatically the measurements. This effect was included in the 
theoretical description of the experiment as direct processes by adding a 
constant to the scattering matrix. The only relevant parameters, whose values 
were taken from the experiment, are the displacement due to the electromagnetic 
impedance and the optical scattering matrix of the elastic waves. An excellent 
agreement between the theoretical predictions and the experimental results was 
obtained.

This work was supported by DGAPA-UNAM under project PAPIIT IN103115. MCS and
AMMA thank the financial support provided by CONACyT. MMM is grateful to the Sistema
Nacional de Investigadores and MA Torres-Segura for her encouragement.

\end{document}